\documentclass[aps,prx,twocolumn,showpacs,superscriptaddress,amsmath,amssymb,floatfix,altaffillsymbol]{revtex4-1}
\usepackage{graphicx}
\usepackage{color}
\usepackage{amsmath}
\usepackage{amssymb}
\usepackage[utf8]{inputenc}
\usepackage{comment}
\usepackage{dcolumn}
\usepackage{bm}
\usepackage{hyperref}
\usepackage{soul}
\usepackage{footmisc}
\usepackage{xfrac}

\newcommand{\Rmnum}[1]{\expandafter\@slowromancap\romannumeral  #1@}

\usepackage{bbold}
\usepackage{color}

\graphicspath{{pics/}}
\begin{document}

\title{Molecular junctions and molecular motors: Including Coulomb repulsion
in electronic friction using nonequilibrium Green’s functions}

\author{M. Hopjan}
\affiliation{Department of Physics, Division of Mathematical Physics, Lund University, 22100  Lund, Sweden}
\affiliation{European Theoretical Spectroscopy Facility, ETSF}

\author{G. Stefanucci} 
\affiliation{Dipartimento di Fisica,Universit\`a di Roma Tor Vergata, Via della Ricerca Scientifica 1, 00133 Rome, Italy}
\affiliation{INFN, Sezione di Roma Tor Vergata, Via della Ricerca Scientifica 1, 00133 Roma, Italy}
\affiliation{European Theoretical Spectroscopy Facility, ETSF}

\author{E.  Perfetto} 
\affiliation{CNR-ISM, Division of Ultrafast Processes in Materials (FLASHit),
Area della Ricerca di Roma 1, Via Salaria Km 29.3, I-00016 Monterotondo Scalo, Italy}
\affiliation{Dipartimento di Fisica,Universit\`a di Roma Tor Vergata, Via della Ricerca Scientifica 1, 00133 Rome, Italy}
\affiliation{European Theoretical Spectroscopy Facility, ETSF}

\author{C. Verdozzi}
\affiliation{Department of Physics, Division of Mathematical Physics, Lund University, 22100  Lund, Sweden}
\affiliation{European Theoretical Spectroscopy Facility, ETSF}

\begin{abstract}
We present a theory of molecular motors based on the Ehrenfest  
dynamics for the nuclear coordinates and the adiabatic limit of the 
Kadanoff-Baym equations for the current-induced forces.
Electron-electron interactions can be systematically 
included  through many-body perturbation theory,  making the 
nonequilibrium Green’s functions formulation suitable for first-principles treatments of realistic  
junctions. The method is benchmarked against 
simulations via real-time Kadanoff-Baym equations, finding an excellent 
agreement. Results on a paradigmatic model of molecular motor show 
that correlations can change dramatically the physical scenario by, 
e.g., introducing a sizable damping in the self-sustained van der Pol 
oscillations.      
\end{abstract}

\maketitle
  \section{Introduction}\label{Introduction}
Ions in a conducting interconnect can drift away from their equilibrium position
due to current-induced forces \cite{Landauer1,Friedel,Black}.
This fact degrades technological performance via, e.g., heating and
electromigration in semiconductor integrated circuits \cite{Blech} 
and nanowires \cite{nanomigration}. However, 
as first envisioned by Sorbello [\onlinecite{Sorbello}], 
current-induced forces can also be turned to one's advantage, with the 
electrons-to-nuclei energy transfer used to move atoms in orbits 
(molecular motors)  
and with prospects of high payoffs for nanotechnology.

{The envision of nanoscale devices converting electrical 
current into mechanical work is attracting a growing interest.}
After the proposal in Ref.~\cite{Sorbello}, a number of
theoretical investigations emerged in 
steady-state  \cite{Todorov2001,Emberly2001,DiVentra2002,Brandbyge2003,Cizek2004,Cornaglia2004,Frederiksen2004,Paulsson2005,
Frederiksen2007,Galperin2007,Galperin2008,Hartle2009,Zhang2011} 
and real-time \cite{Horsfield2004a,Horsfield2004b,Verdozzi2006,Sanchez2006,
Todorovic2011,Albrecht2012} transport to understand and possibly manipulate current-induced forces.
Their nonconservative character was pointed out in several studies 
\cite{DiVentra2004,Dundas2009,Todorov2010}. It was also pointed out 
that these forces are of two types, i.e., 
friction-like~\cite{Hussein2010,Lu2011,Lu2012} and Lorentz-like~\cite{Lu2012,Lu2010}. 
Under general nonequilibrium conditions the friction force
can be negative and responsible for van der Pol oscillations of the nuclear
coordinates \cite{Bennett2006,Bode2011,Bode2012,Hussein2010}, runaway modes \cite{Verdozzi2006,Lu2010} or heating \cite{Lu2015}.

Interestingly enough, electronic correlations in these situations 
(and thus in concept-protocols of molecular motors) have not been 
addressed until very recently.    
A first step was taken by Dou et al. 
\cite{Dou2017a}, with a general formulation in terms of 
{\em $N$-particle} Green's functions, $N$ being the number 
of electrons in the system (see also \cite{Dou2018,Chen2018}
for subsequent discussions).
Afterwards, an expression for the friction force 
was derived via a generalized master equation in the Coulomb blockade 
regime~\cite{Calvo2017}. 

A fundamental merit of these two pioneering works is to bring the 
issue of electronic correlations in molecular motors into the 
spotlight. However, it is also the case that, at present, a general approach 
suitable for 
calculations of nuclear motion in realistic junctions is still lacking.  Also, 
an assessment of the importance of second- and 
higher-order corrections in the 
nuclear velocities of the current-induced forces~\cite{Verdozzi2006,Metelmann2011,Nocera2011,Kartsev2014}
has not yet been made.

Motivated by these considerations, we derive here a 
formula of current-induced forces in terms of the 
{\em one-particle} steady-state nonequilibrium Green's function (ssGF).
The main advantage of the ssGF formulation is that electronic correlations 
can be systematically and self-consistently included through 
diagrammatic approximations to the many-body self-energy, 
particularly suitable in first-principle approaches.
Like previous non-interacting formulations, we account only 
for the lowest order correction in the nuclear velocities. 
The impact of higher-order corrections is assessed through 
benchmarks against mixed quantum-classical studies based on 
Ehrenfest dynamics (ED) for the nuclei and either the two-times 
Kadanoff-Baym equations~\cite{KBE,Keldysh,StefLeeu,BalzBon,Hopjan14,commentKBE,Bostrom2016,Balzer2016}   
(KBE) or the one-time Generalized Kadanoff-Baym Ansatz~\cite{Lipavsky1986} (GKBA) for the 
electronic part. We find that the ssGF scheme is quantitatively accurate 
and numerically highly efficient. The main physical result of our 
investigations is that electronic correlations hinder the emergence 
of negative friction.

\section{Nonadiabatic Ehrenfest Dynamics}\label{NED}
  
We consider a metal-device-metal junction and a set of classical 
nuclear coordinates ${\bf x}=\{x_{1},x_{2},\ldots\}$  coupled  
to electrons in the device. 
The junction is exposed to time-dependent 
gate voltages  and  biases. 
For heavy nuclear masses ${\bf M}=\{M_{1},M_{2},\dots\}$ an expansion 
of the nuclear wave functions around the classical trajectories  
yields~\cite{Hussein2010,Metelmann2011} ($T$ labels time)
\begin{align}\label{lange}
M_{\nu} d^{2}x_{\nu}/dT^{2}=-\partial_{x_{\nu}}\mathcal{U}_{\rm cl.}({\bf x}(T))+
F^{\rm el.}_{\nu}[{\bf x},T]-\xi_{\nu} ,
\end{align}
where $\mathcal{U}_{\rm cl.}({\bf x})$ is the classical potential 
of the nuclei, $F^{\rm el.}_{\nu}[{\bf x},T]$ is the   
force exerted by the electrons and $\xi_{\nu}$ is a stochastic contribution 
\cite{Hussein2010,Metelmann2011,Bode2011,Bode2012,Lu2012}. 
For $\xi_{\nu}=0$ the Langevin-type Eq.~(\ref{lange})
reduces to the ED equation. In the following the stochastic field
will be neglected.

The most general device Hamiltonian can be written as
\begin{align}
H_{\rm C}({\bf x},T)=\sum_{ij,\sigma}h^{}_{ij}({\bf 
x},T)c^{\dagger}_{i\sigma}c^{}_{j\sigma}+H_{\rm int.},
\label{Hcentrale}   
\end{align}
where $c^{\dagger}_{i\sigma}$ creates an electron with spin 
projection $\sigma$ on the $i$-th localized orbital of the device 
region. The term $H_{\rm int.}$ is independent of ${\bf x}$ and 
accounts for electron-electron interactions. The electronic force 
then reads 
\begin{eqnarray}
\label{hamilt1}
F^{\rm el.}_{\nu}[{\bf x}(T),T]&=&- \left.\langle \partial_{x_{\nu}} H_{\rm C}({\bf x},T) \rangle
\right|_{{\bf x}={\bf x}(T)}
\nonumber \\
&=&-\sum_{ij,\sigma}  \rho_{ji}(T) \partial_{x_{\nu}} h_{ij}({\bf x}(T),T).
\label{elforce}
\end{eqnarray}
where $\rho_{ji}(T)$ is the electronic one-particle density matrix.
In general, $\rho$ depends on the history of the system, and so does 
the electronic force via $\rho$, as evident from Eq.~\eqref{elforce}. This is generally
referred to as "non-Markovian dynamics".
Below we discuss two ways how to perform the time evolution of the electronic
density matrix which includes the memory effects, both formulated
in the nonequilibrium Green's functions (NEGF) framework.

\subsection{Kadanoff-Baym equations}\label{KBE}

In the NEGF formalism \cite{BalzBon,StefLeeu,Hopjan14}, the density matrix 
$\rho_{}$ can be calculated from the equal-time lesser Green's function 
according to $\rho_{}(T)= -i G^{<}_{}(T,T^{+})$. The double-time lesser Green's function $G^{<}(t,t')$ is obtained from the contour Green's function 
$G(z,z')$
by setting $z=t$ on the forward branch and $z'=t'$ on the backward 
branch of the Keldysh contour 
$\gamma$~\cite{KBE,Keldysh,BalzBon,StefLeeu,Hopjan14,commentKBE}. 
The contour time evolution is governed by the equation of motion~\cite{StefLeeu,BalzBon,Hopjan14} 
\begin{align}
\label{kbe}
&[~{\rm i}\partial_{z}-h_{\rm HF}({\bf x}(z),z)]~G(z,z')= \nonumber
\\&~~~~~~=\delta(z,z') \mathbb{1}
 +\int_{\gamma}({\Sigma}_{\rm corr.}+{\Sigma}_{\rm 
 emb.})(z,\bar{z})G(\bar{z},z')d\bar{z}
\end{align}
where $h_{\rm HF}=h_{}+{\Sigma}_{\rm HF}$ is the sum of the 
single-particle Hamiltonian and  Hartree-Fock (HF) self-energy.
The self-energy ${\Sigma}_{\rm corr.}$ accounts for electronic 
correlations beyond Hartree-Fock whereas ${\Sigma}_{\rm emb.}$ is the 
standard embedding self-energy.
A similar equation holds for $z'$. 
Choosing  $z$ and $z'$ on different 
branches and breaking the contour
integral into real-time integrals one obtains the KBE, 
see Supplemental Material (SM) \cite{SM}. They are
coupled to the nuclear ED through Eq.~(\ref{lange}) resulting in a scheme
that in the following we refer to as
ED+KBE.  In this work 
we solve the ED+KBE using the Second Born approximation (2BA) to 
${\Sigma}_{\rm corr.}$~\cite{conserv}, 
whose performance has been tested previously (see, e.g., Ref.~\cite{Puig2009,commentKBE}).
The physical picture behind the 2BA is that two electrons, in addition to feel a mean-field generated by all other
electrons, can also scatter directly once (see also the SM).

\subsection{Generalized Kadanoff-Baym Ansatz}\label{GKBA}  

The KBE scale as $N_T^3$, $N_T$ being the time grid size \cite{Bonitzpaper}.
To reduce memory costs, the time propagation can be directly
performed for $\rho$. Formally, the general exact equation for $\rho$ can be derived from the KBE at 
equal times, i.e. on the time diagonal $t=t'$: 
\begin{align}
\frac{d\rho(t)}{dt}+{\rm i}[h_{\rm HF}({\bf x}(t),t),\rho(t)]=-(I(t)+{\rm H.c.}),
\label{rhoeom}
\end{align}
where the collision integral $I$ involves lesser (denoted by "$<$") and greater (denoted by "$>$")
components of the two-times functions 
$G$, ${\Sigma}_{\rm corr.}$ and ${\Sigma}_{\rm emb.}$
To close the equation for $\rho$
we make the Generalized Kadanoff-Baym Ansatz~\cite{Lipavsky1986} 
\begin{eqnarray}
&G^{<}(t,t')=-G^{R}(t,t')\rho(t')+\rho(t)G^{A}(t,t'),
\end{eqnarray}
where a specification for $G^{R/A}$ is needed which, in this paper, 
is made in terms of  the so called static-correlation approximation ~\cite{Latini2014} (see also SM for details). 
When combining the GKBA with the ED (henceforth referred to as ED+GKBA), we 
use the 2BA for ${\Sigma}_{\rm corr.}$, consistently
with the ED+KBE scheme discussed above. For purely electronic dynamics, 
the two schemes were shown to be in good mutual agreement ~\cite{Latini2014},
especially for not too strong interactions. Finally, 
one-time ED+GKBA evolution allows for much longer propagations
than the two-time ED+KBE scheme.

\section{Adiabatic Ehrenfest Dynamics}\label{AED}

 As discussed above, in general
electrons and nuclei obey coupled equations of motion
(Eqs.~(\ref{lange}) and  (\ref{kbe}), or Eqs.~(\ref{lange}) and (\ref{rhoeom})),
and memory effects should be taken into account in the electron dynamics. 
In this section we show that, under specific assumptions, a simplification occurs,
namely for slow nuclear dynamics the equations can be decoupled and one can
propagate only Eq. \eqref{lange}. 

If the nuclear velocities $\dot{\bf x}$ are small, the
electronic force can be expanded up to linear order in nuclear velocities
$ \dot{\bf x}$. Additionally, in the adiabatic limit where the memory effect are
negligible, the coefficients of the expansion can be determined by the electronic
steady state corresponding to the fixed nuclear position $\bf x$ (also known as
Markovian or nonequilibrium Born-Oppenheimer assumption). Under these conditions
the electronic force can be divided into two contributions $F^{\rm el.}\approx F^{\rm ss}_{}[{\bf x}]
+F^{\rm fric}_{}[{\bf x},{\dot{\bf x}}]$ where the first term is the
steady-state force and the second one is the friction+Lorentz like force.
These forces, known as current-induced forces, are introduced below
in terms of the {\em one-particle} steady-state nonequilibrium Green's function (ssGF).

\subsection{Current induced forces}\label{ssKBE}  

At the steady state, where ${\bf x}$ is time-independent, one can find the corresponding 
steady-state Green's functions ${G}^{}_{\rm ss}$ containing information about densities and currents
in the system. The Green's functions depend only on the frequency
$\omega$ and satisfy the steady-state KBE ({in matrix form and} omitting the
parametric dependence on ${\bf x}$):
\begin{align}\label{gadi}
{G}^{R}_{\rm ss}(\omega)&= \frac{1}{\omega-{h}_{\rm HF}-{\rm \Sigma}^{ 
R}_{\rm ss}(\omega)} \nonumber \\
{G}^{<}_{\rm ss}(\omega)&= \frac{1}{\omega-{h}_{\rm HF}-{\rm \Sigma}^{ 
R}_{\rm ss}(\omega)}
{\rm {\rm \Sigma}^{<}_{\rm ss}(\omega)
\frac{1}{\omega-{h}_{\rm HF}-{\rm 
\Sigma}^{A}_{\rm ss}(\omega)}},
\end{align}
where $\Sigma_{\rm ss}$ is the steady-state value of ${\Sigma}={\Sigma}_{\rm corr.}
+{\Sigma}_{\rm emb.}$. The lesser steady-state Green's function ${G}^{<}_{\rm ss}$ gives direct
access to the steady-state force,
\begin{equation}
F^{\rm ss}_{\nu}[{\bf x}]={+} 2{\rm i}\int \frac{d\omega}{2\pi} 
{\rm Tr}\left[G^{<}_{\rm ss}[{\bf x}](\omega)\partial_{x_{\nu}}h({\bf 
x})\right],
\label{ssforce}
\end{equation}
while the friction$+$Lorentz-like is obtained as
\begin{eqnarray}
F^{\rm fric}_{\nu}[{\bf x},\dot{\bf x}]&=
& {-}\sum_{\mu}\dot{x}_{\mu}\gamma_{\nu\mu}
{[{\bf x}]}.
\label{fricforce}
\end{eqnarray}
Here the friction coefficients $\gamma_{\nu\mu}$ are 
dependent on the parameter ${\bf x}$ through ${G}^{}_{\rm ss}$. Explicitly,
\begin{align}\label{frictioneq}
\gamma_{\nu\mu}[{\bf x}]=\int \frac{d\omega}{2\pi}{\rm Tr}
\Bigl[
\Bigl(&{\mathcal Q}_\mu({G^{ R}_{\rm ss}},h_{\rm HF}+\Sigma^{R}_{\rm ss,corr.},{G^{\rm <}_{\rm ss}})
\nonumber \\
+ &{\mathcal Q}_\mu({G^{\rm <}_{\rm ss}},h_{\rm HF}+\Sigma^{A}_{\rm ss,corr.},{G^{ A}_{\rm ss}})
\nonumber \\
+ &{\mathcal Q}_\mu({G^R_{\rm ss}},{\rm \Sigma^<_{\rm ss, corr.}},{G^A_{\rm ss}})\Bigr)(\partial_{x_{\nu}}{h})\Bigr],
\end{align}
where ${\mathcal Q}_\mu(a,b,c)= [ (\partial_\omega a) 
(\partial_{x_{\mu}}b)  c - a  (\partial_{x_{\mu}} b) (\partial_\omega 
c) ]$. 

The result in Eq.~\eqref{frictioneq} applies to systems with electron-electron
interactions and provides an alternative to the friction formula in terms of
{\em $N$-particle} Green’s functions \cite{Dou2017a,Proof}. Furthermore,
Eq.~\eqref{frictioneq} directly reduces to previously published results in the 
noninteracting case \cite{Bode2011,Bode2012,Equiv}. 
More important, the advantage of the presented expression for the 
friction force is that electronic correlations can be systematically and 
self-consistently included through diagrammatic approximations \cite{Remark}.

\subsection{Derivation of current induced forces from KBE}\label{ssKBE}  

The current induced forces presented above can be derived from the nonadiabatic
KBE dynamics in the adiabatic limit. In the following we briefly discuss the 
main steps of the derivation (for the full derivation see the SM):

{\em i)} We start with the nonadiabatic KBE dynamics where the electronic
evolution is characterized by the two-times Green's functions $G(t,t')$
and we move to the Wigner representation $G(t,t') \rightarrow 
{G}(\omega, T)$ \cite{Wigner}, where $T = \frac{t+t'}{2}$ is the center-of-mass time
and $\omega$ is the Fourier conjugate of the relative time $\tau=t-t'$.

{\em ii)} Under the assumption that the nuclear velocities are small we can
expand $G^{<}$ and $G^{R}$ in powers of the 
nuclear velocities $\dot{\bf x}$ \cite{Moyal}. To first order one finds 
$G^{<}(\omega,T)={G}^{<}_{\rm ss}(\omega)+
{\rm i}\sum_{\mu}\dot{x}_{\mu}(T)\Delta_{\mu}(\omega,T)$
where $\Delta_{\mu}$ is a complicated function of $G^{<}$, 
$G^{R}$ and their derivatives with respect to $\omega$ and 
$x_{\mu}$. This expansion consistently preserves the general
relation $G^{>}-G^{<}=G^{R}-G^{A}$ for any finite bias~\cite{botermans}.

{\em iii)} Subsequently, we invoke the assumption of adiabatic (Markovian) limit. We evaluate
$\Delta_{\mu}$ at the steady-state Green's functions, thus
obtaining $\Delta_{\mu}(\omega,T)\rightarrow\Delta_{\mu,\rm ss}(\omega)$. Then, we take into 
account that $ \rho(T)=-{\rm i}\int \frac{d\omega}{2\pi} 
{G}^{<}(\omega,T)$ in Eq.~(\ref{elforce}). The integral gives access to  the steady 
state force and the friction$+$Lorentz-like force.

As the nonadiabatic dynamics (ED+KBE or ED+GKBA) is the starting point to derive the adiabatic
dynamics (ED+ssGF), the former can be used to benchmark the latter in the adiabatic limit. 

The advantage of the ED+ssGF scheme is in its computational efficiency. Once the values
of the steady state and friction force are computed and tabulated (for each $\bf x$), one can
evolve the nuclear
coordinates {\em for any initial condition} using only Eq. \eqref{lange}.

\section{Dynamics of Model System}\label{system}

\begin{figure}[t!]
\begin{center}
\includegraphics[width=8.5cm]{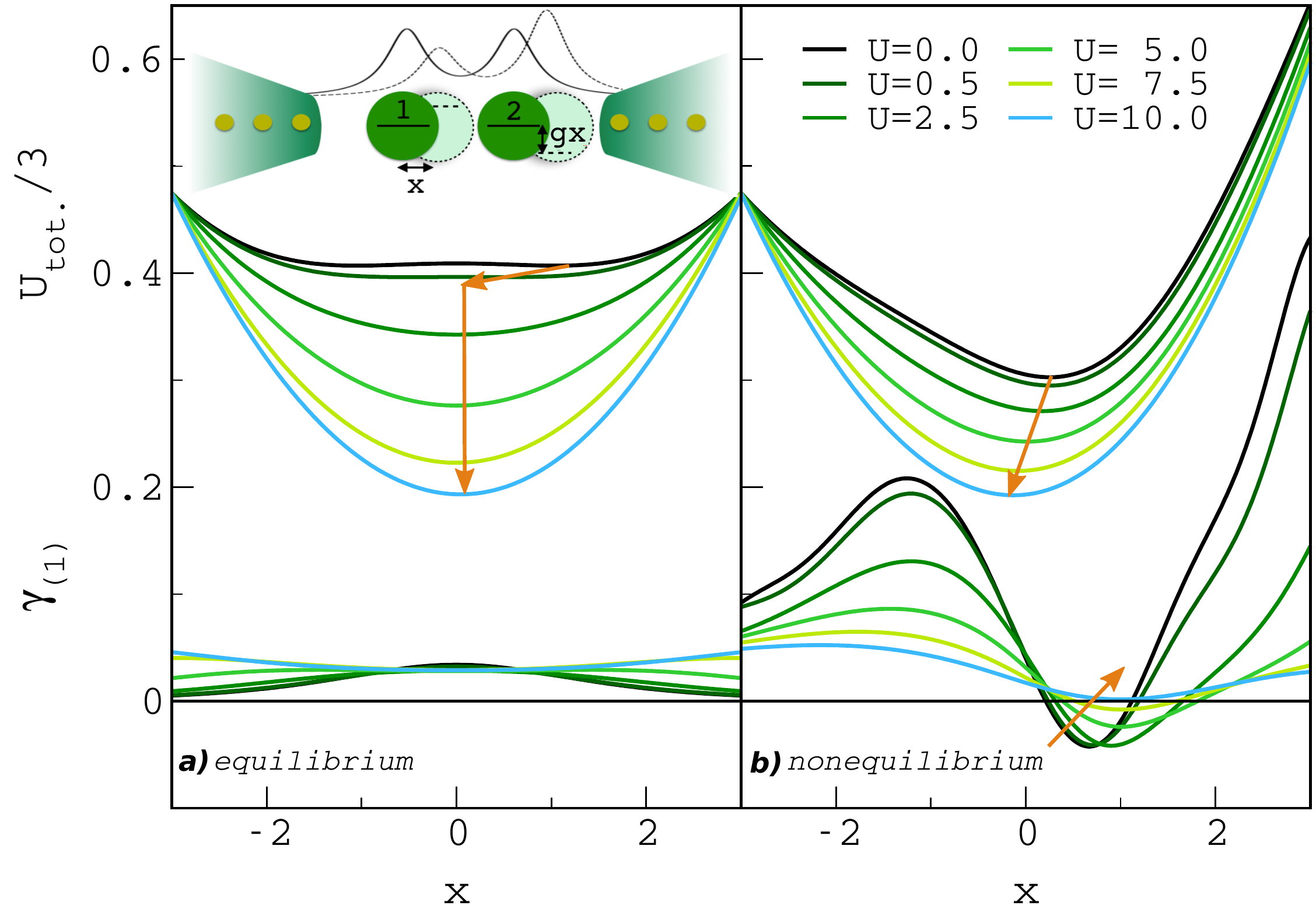}
\caption{{Total potential $\mathcal{U}_{\rm tot.}$ and friction 
$\gamma$ in 2BA as function of $x$ for different
interaction strengths $U$ in (a) equilibrium with $v_{c}=0$ and (b)
at finite bias $V_{L}=-V_{R}=5$ with $v_{c}=1$. 
The system parameters are:  
$g=1.58$, ${J}_{c}=-3.5$, $J=50$ and ${J}_{\rm tun.}=-8.66$.
The inset shows a dimer (green circles) coupled to leads, the effective energy of 
sites 1 and 2 (horizontal tracts) and the charge density (lines over 
the dimer) for 
$x=0$ (solid lines) and for $x>0$ (dashed lines).}} 
\label{friction}
\end{center}
\vspace{-0.8cm}
\end{figure}

We demonstrate the impact of electronic 
correlations in the model system originally introduced in 
Ref.~\cite{Hussein2010}, namely a dimer { 
that can rigidly oscillate with frequency $\Omega$ between two leads, see 
inset in Fig.~\ref{friction}-a)}.
As in Ref.~\cite{Hussein2010}, {we 
express all energies in units of $\hbar\Omega$,
times in units of $1/\Omega$ and distances in units 
of the characteristic harmonic oscillator length  
${l_{0}}=\sqrt{\hbar/(M\Omega)}$.}
The dimensionless dimer Hamiltonian 
reads
\begin{eqnarray}
H_{\rm 
C}(x,T)&=&\sum_{\sigma}{J}_{c}(c_{1\sigma}^{\dagger}c_{2\sigma}^{}+{\rm H.c.})+
{v}_{c}(T)\sum_{i\sigma}n_{i\sigma}
\nonumber\\
&+&gx\sum_{\sigma}(n_{1\sigma}-n_{2\sigma})+U\sum_{i}
n_{i\uparrow}n_{i\downarrow},
\end{eqnarray}
where we added a Hubbard-like interaction (last term) to the original 
model. { The electron-nuclear coupling has strength $g$ 
and describes a dipole-dipole interaction.}
The dimer is {further} connected to a left (L) lead through site 1 and 
to a right (R) lead through site 2 with hopping amplitude 
$J_{\rm tun.}$. The L/R lead is a semi-infinite
tight-binding chain with 
nearest neighbor hopping integral $J$ and {time-dependent} onsite energy (bias) $V_{\rm 
L/R}(T)$.\\

In Fig.~\ref{friction} we plot the total potential $\mathcal{U}_{\rm 
tot.}=\mathcal{U}_{\rm cl.}+\mathcal{U}_{\rm ss}$ where  
$\mathcal{U}^{}_{\rm ss}=-\int^{x}_{-\infty}F^{\rm ss}dx$
and  friction coefficient $\gamma=\gamma_{11}$, both 
calculated within the 2BA \cite{conserv}.   
In equilibrium, hence $V_{\rm L/R}=0$,  the system is symmetric under the 
inversion of $x$ and so are potential and friction, {see 
panel a)}. 
{For $U=0$ we have a double minimum in 
$\mathcal{U}_{\rm tot.}$ corresponding to the two degenerate 
Peierls-distorted ground states. With increasing $U$ the 
repulsive-energy cost of the charge-unbalanced Peierls states
becomes larger than the distortion-energy gain. Consequently, 
$\mathcal{U}_{\rm tot.}$ develops a single minimum in $x=0$ and the 
charge-balanced ground state becomes favored.}
{Independently of $U$,  the friction remains 
positive, an exact equilibrium property 
correctly captured by our diagrammatic 2BA.}

Turning on a 
gate voltage $v_{c}=1$ and a bias $V_{\rm L}=-V_{\rm R}=5$,
{see panel b),}
{electrons start flowing through the dimer}.
The noninteracting formulation predicts self-sustained van der Pol oscillations 
\cite{Bode2011,Bode2012,Kartsev2014} since the minimum in  $\mathcal{U}_{\rm 
tot.}$ occurs for values of $x$ where $\gamma$ is negative.
{Thus, the electrical current activates an ever lasting
sloshing motion of the dimer.}
Electron correlations
shift  the position of the potential minimum
away from the region $\gamma<0$, thus hindering 
the van der Pol oscillations.
This effect is even enhanced by the flattening of $\gamma$ that 
causes a shrinking of  the region of negative friction.
We point out that the HF approximation, i.e., $\Sigma_{\rm corr.}=0$, 
predicts the opposite behavior. 
To validate the correctness of the 2BA treatment
we have evaluated $\gamma$ also within the T-matrix approximation \cite{conserv}
(TMA), which accounts for multiple scattering of electrons, and found similar results (see SM).\\

\begin{figure}[t!]
\begin{center}
\includegraphics[width=8.5cm]{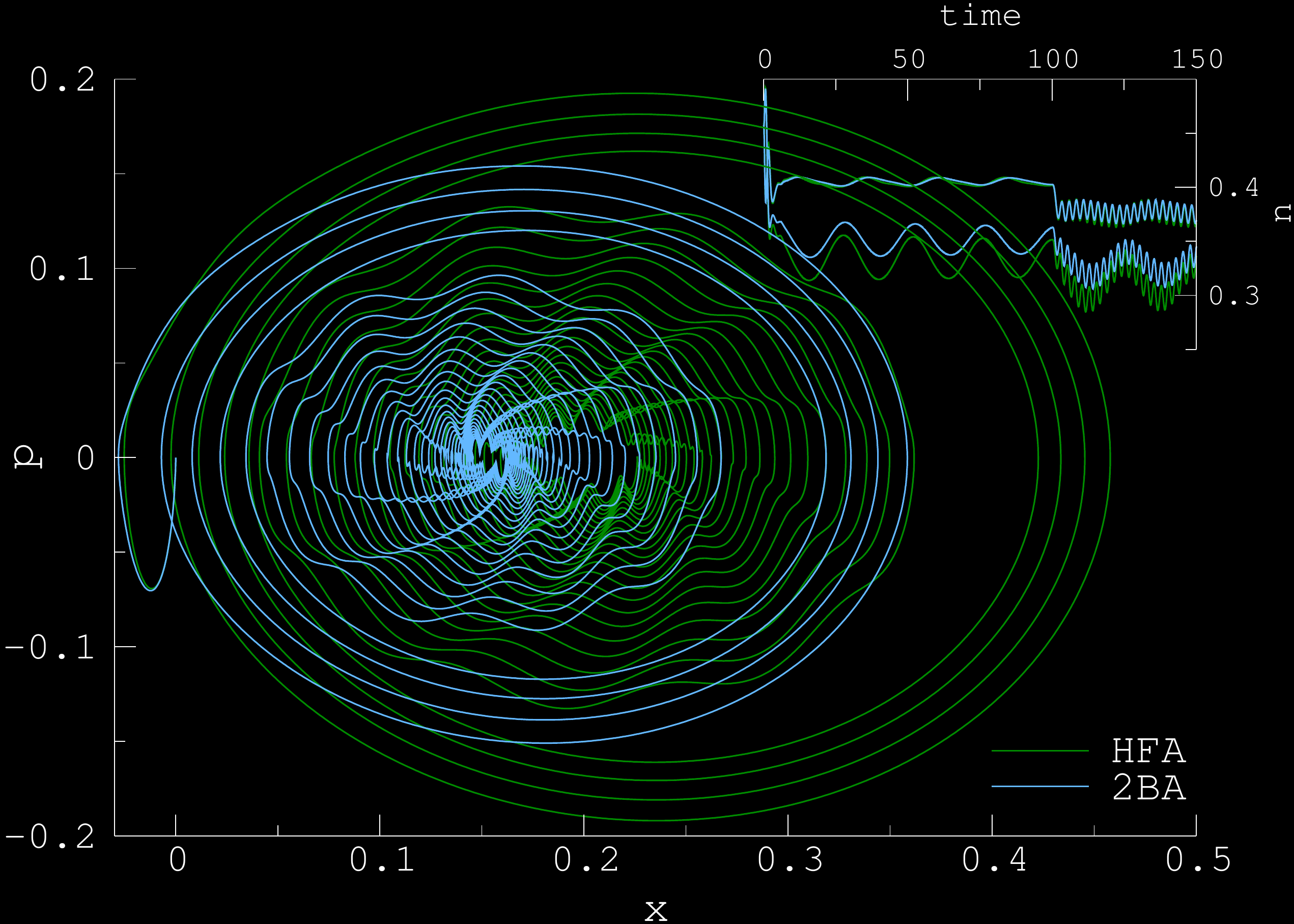}
\caption{{Phase space $(p,x)$ trajectories in ED+GKBA with gate  
$v_{c}(T)=\theta(T)[1+{\sin}^2(\frac{2\pi}{5} 2gT)]$ 
and $U=2.5$ in the HF and 2B approximations (time rescaled by 
$1/(2g)$).
The inset (top right) shows the density of the two sites of the 
dimer.}}
\label{nonadiabatic}
\end{center}
\vspace{-0.8cm}
\end{figure}

The differences between the HF and 2BA results are illustrated in 
time domain in Fig.~\ref{nonadiabatic} using the ED+GKBA approach. 
We start from an equilibrium situation and then switch on a bias 
$V_{\rm L}=-V_{\rm R}=5$ and a gate $v_{c}=1$.  Then,
after time $t=100$, we add a high-frequency time-dependent gate whose
only effect is to modulate the nuclear trajectory; ultrafast field 
{have only a minor influence in steering  molecular motors.
Notice that,} although 
$U/J_{c}\approx 0.7$ (weakly correlated regime), the HF and 2BA 
trajectories are quantitatively very different.\\

\begin{figure}[h!]
\begin{center}
\includegraphics[width=8.6cm]{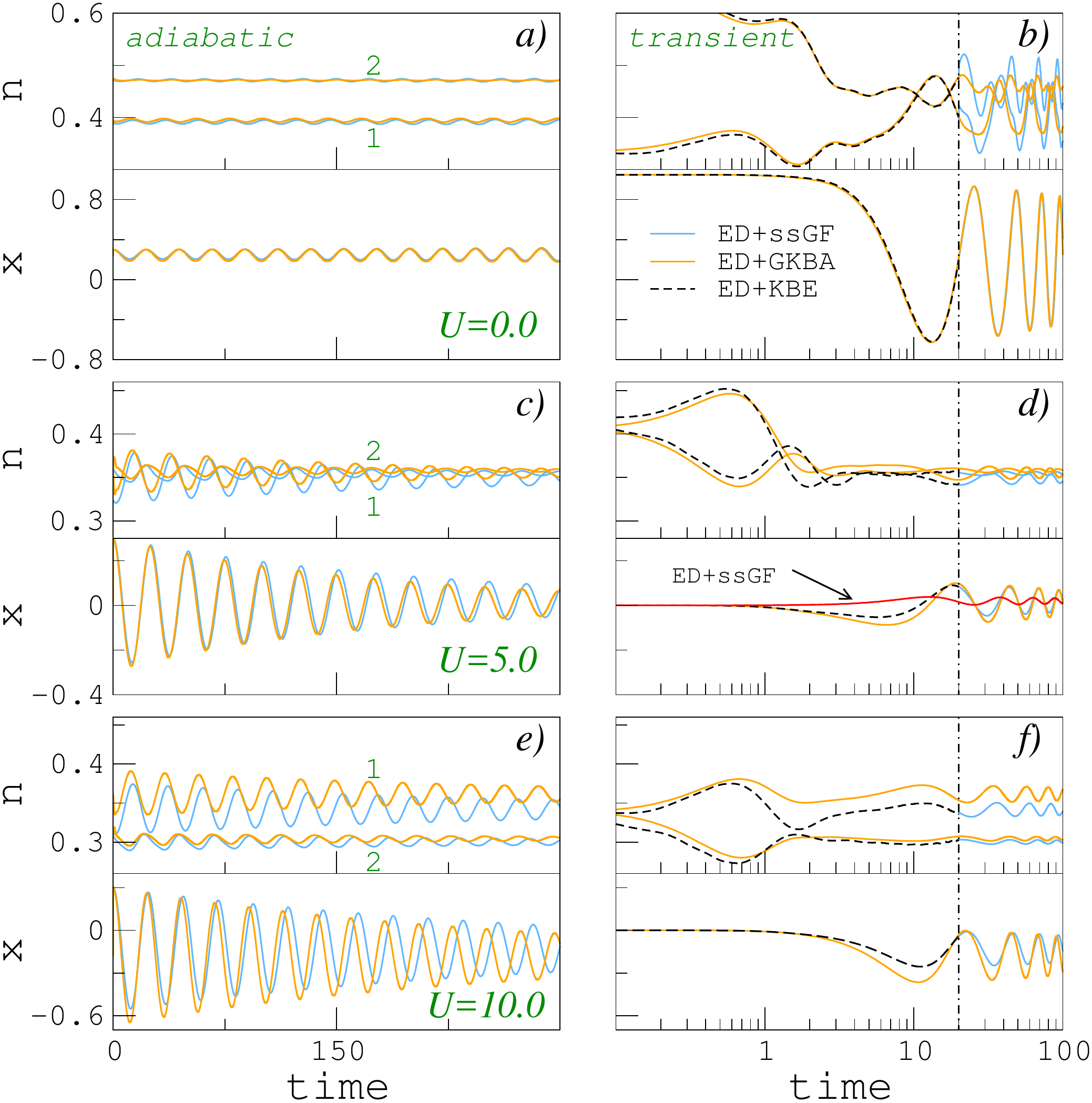}
\caption{Comparison between ED+GKBA and ED+ssGF for  nuclear coordinate
and dimer densities in the 2B approximation
(time rescaled by $1/(2g)$).} 
\label{adiabatic}
\end{center}
\vspace{-0.8cm}
\end{figure}

{The effects of Coulomb interactions on the electromechanical 
energy conversion is investigated in Fig.~\ref{adiabatic}.}
In the left panels 
we consider a steady-state system with a bias
at time $t=0$ and then 
suddenly change the position of the nuclear coordinate to 
$x=0.3$. No external fields other than the bias are switched on, so all quantities 
depend on time {only} through $x$. 
{Simulations are performed with ED+ssGF and ED+GKBA 
(the maximum propagation time is too long for ED+KBE)}.  
The nuclear coordinate and site densities show an excellent agreement between the 
two schemes up to $U=5$.  
The real-time simulations confirm the 
conclusions drawn by inspection of Fig.~\ref{friction}. 
{van der 
Pol oscillations are ever lasting only for $U=0$ [panel a)]; 
for $U>0$ the dynamics is damped [panels c)-e)]. We can estimate the size of the 
effect for a normal mode with period $T=2\pi/\Omega\simeq 10^{1}$~fs 
(hence $V_{L}-V_{R}= 10\hbar\Omega\simeq 1\div 10$~eV). 
In this case the average current through the dimer is in the $\mu$A 
range (which is congruous for molecular transport) and
the amplitude of the sloshing motion is, from Fig.~\ref{adiabatic}, 
of the order $l_{0}\simeq  10^{-1}\div 10^{-2}~{\rm \AA}$
(we assumed a dimer of mass $M\sim 25 M_{\rm proton}$ which is appropriate for molecules like, e.g., 
ethylen). Then for $U=5\hbar\Omega\simeq (0.1\div 1)$~eV the 
Coulomb-induced damping occurs on the picosecond timescale (see also SM for details).}

In the right panels of Fig.~\ref{adiabatic} 
we explore the performance of the ED+ssGF scheme 
for a situation when the system has not yet attained a steady state.
At time $t=0$ we switch-on a constant (in time) gate $v_{c}=1$ 
and bias $V_{\rm L}=-V_{\rm R}=5$ and propagate the system 
using both ED+KBE and ED+GKBA. After a transient phase [time window 
$(0,20)$] we continue the ED+KBE propagation using the ssGF scheme with 
initial condition given by the ED+KBE value of the nuclear coordinate 
at time $t=20$. The duration of the transient phase was chosen longer 
than the tunneling time in order to wash out the effects of the 
sudden switch-on of the external fields. 

In the noninteracting case, panel b), the system is in a strong 
nonadiabatic regime, and the ssGF densities are largely  deviating 
from the GKBA densities, especially close to the maxima of $|\dot{x}|$. 
Nevertheless, the 
ssGF and GKBA nuclear coordinates are
almost identical. This is a consequence of 
the fact that also the density deviations on the two sites are almost 
identical, and hence the electronic force (which depends on the 
densities difference) is not affected by these deviations.
 
For $U=5$, panel d), the system is in the adiabatic regime 
after the transient, and we observe a good agreement between the ED+GKBA 
and the ED+ssGF dynamics. To appreciate the importance of 
nonadiabatic effects  we also 
plot the result of the pure ED+ssGF
dynamics (red line). During the transient ED+ssGF is not expected to 
work since we are not close to the KBE steady state. Interestingly, 
however, the impact of the sudden switch-on  is strong also at long 
times; the ED+ssGF nuclear coordinate disagrees 
considerably from that of ED+GKBA.    
Increasing the interaction further, panel f), the ED+GKBA dynamics 
starts to deviate from the ED+KBE dynamics, with a sizable 
overestimation of the amplitude of the oscillations. This is again a 
consequence of the failure of the GKBA for too strong $U$'s.
  
  \section{Conclusions}\label{conc}

We introduced a theoretical description of molecular motors in
molecular junctions, based on a coupled quantum-classical approach,   
with nuclei treated within the Ehrenfest dynamics (ED), and electrons
within the two-times Kadanoff-Baym Equations (KBE) or the one-time 
Generalized Kadanoff-Baym Ansatz (GKBA).

In the adiabatic limit of these descriptions, we   used the steady-state
nonequilibrium Green's function (ssGF) to derive an expression for the
electronic friction coefficient which includes correlation effects due to Coulomb repulsions among the electrons.
The adiabatic assumption allows for integrating out the electronic 
degrees of freedom thus providing a description of the nuclear 
dynamics in terms of forces that can be calculated and stored in 
advance.  We demonstrated that the
proposed ED+ssGF approach is accurate and computationally more
efficient than ED+KBE and even ED+GKBA.

We considered the paradigmatic Hubbard dimer to investigate 
the role of correlations and performed calculations in the mean-field HF 
approximation as well as in the correlated 2BA 
and TMA {to treat the Coulomb interaction}. 
Numerical evidence indicates that {the HF approximation is}
not accurate enough and that correlation effects can change 
dramatically the 
physical picture. In fact, in a broad range of model parameters we found that
correlations hinder the emergence of regions
of negative friction and strongly damp the nuclear motion.
Our results also suggest that fast driving 
fields play a minor role in designing molecular motors. 

Of course, the investigation of electronic correlations in molecular 
motors is still at its infancy. The proposed ED+ssGF approach 
allows for standard diagrammatic approximations and therefore well suited 
for first-principle treatments of realistic setups. We envisage its 
use to gain insight into molecular devices, 
and hopefully  to put technological applications at a closer reach.

\begin{acknowledgements} 
We acknowledge D. Karlsson for discussions and E. Bostr\"om for critically reading the manuscript.
E.P. acknowledges funding from the European Union project MaX 
Materials design at the eXascale H2020-EINFRA-2015-1, Grant Agreement 
No. 676598 and Nanoscience Foundries and Fine Analysis-Europe 
H2020-INFRAIA-2014-2015, Grant Agreement No. 654360.   
G.S. acknowledges funding by MIUR FIRB Grant No. RBFR12SW0J and EC funding through the
RISE Co-ExAN (GA644076).
\end{acknowledgements}

\bibliographystyle{andp2012}

\providecommand{\WileyBibTextsc}{}
\let\textsc\WileyBibTextsc
\providecommand{\othercit}{}
\providecommand{\jr}[1]{#1}
\providecommand{\etal}{~et~al.}

\end{document}